\documentclass[preprint2]{emulateapj}

\slugcomment{}

\shorttitle{Weighing the non-transiting hot Jupiter $\tau$ Boo b}
\shortauthors{Rodler et al.}

\begin{document}


\title{Weighing the non-transiting hot Jupiter $\tau$~Boo~b}


\author{F. Rodler\altaffilmark{1}}
\email{rodler@ieec.uab.es}
\author{M. Lopez-Morales\altaffilmark{1,2}}
\author{I. Ribas\altaffilmark{1}}

\altaffiltext{1}{Institut de Ci\`encies de l'Espai (CSIC-IEEC), Campus UAB, Facultat de Ci\`encies, Torre C5 parell, 2a pl, 08193, Bellaterra, Spain}

\altaffiltext{2}{Visiting Investigator. Carnegie Institution of Washington, Department of Terrestrial
Magnetism, 5241 Broad Branch Road NW, Washington,
DC 20015-1305, USA}
\begin{abstract}

We report the detection of the orbital velocity of non-transiting hot Jupiter $\tau$~Boo~b. By employing high-resolution ground-based spectroscopy around 2.3~$\mu$m during one half night, we are able to detect carbon monoxide absorption lines produced in the planet atmosphere, which shift significantly in wavelength during the course of the observations due to the orbital motion of the planet. This detection of the planetary signal results in the determination of the orbital inclination being $i=47 ^{+7}_{-6}$~degrees and furthermore allow us to solve for the exact planetary mass being $m_{\rm p}=5.6 \pm 0.7~\rm{M_{\rm{Jup}}}$. This clearly confirms the planetary nature of the non-transiting companion to $\tau$~Boo. 

\end{abstract}


\keywords{planets and satellites: atmospheres --- planets and satellites: individual ($\tau$~Boo~b) --- techniques: radial velocities --- techniques: spectroscopic}

\section{Introduction}

The radial velocity (RV) technique is the most successful exoplanet detection method, with more than 600 planets\footnote{The Extrasolar Planets Encyclopedia; http://www.exoplanet.eu} detected to date. This technique monitors the variation of the RV of the star due to the gravitational pull of an unseen companion. Owing that the RV of the planet is not measured directly, neither the orbital inclination $i$ of the planet nor its exact mass can be determined. Instead, only an estimation of the minimum mass of the planet can be derived by applying Kepler's Laws, which give $m_{\rm p, min}=m_{\rm p}\sin i$, where $m_{\rm p}$ is the unknown planetary mass. For transiting planets - i.e. for planets which periodically occult their host stars - the orbital inclination $i$ and therefore the exact planetary mass can be measured. However, the vast majority of extrasolar planets found to date are non-transiting and therefore without an exactly determined mass. Consequently, many of these non-transiting planets, especially the ones with largest minimum masses, can only be labelled as  {\it planetary candidates}, owing to the case that for very low orbital inclination they might turn out to be brown dwarfs. 

One strategy to determine the exact mass of non-transiting planets is to directly measure the planetary RV signal via high-resolution (i.e. $R=\lambda / \Delta \lambda > 40,000$; $\lambda$ denotes the wavelength) spectroscopy with very large telescopes. Key to this method is to observe a large number of spectral features coming from the planet and to observe them at different orbital phases so the traveling faint planetary signal can be disentangled from the dominating stellar one. In the past, several high-resolution spectroscopy campaigns at the optical with the goal to detect starlight reflected from hot Jupiters (i.e. massive planets that are a few stellar radii away from their host stars) and to measure their exact masses were carried out. Although all of them resulted in non-detections of reflected light, these campaigns confirmed the low reflectivity of hot Jupiters at visual wavelengths (e.g. \citealp{1999ApJ...522L.145C,2002MNRAS.330..187C,2003MNRAS.344.1271L,2008A&A...485..859R}). Towards near-infrared (NIR) wavelengths, the planet-to-star flux ratios drastically increase due to the strong thermal emission of hot Jupiters. \cite{2001ApJ...546.1068W,2007MNRAS.379.1097B,bar08,bar10} and \cite{cub11} observed hot Jupiters by means of high-resolution spectroscopy at near-infrared wavelengths, but were not able to detect any molecules in their atmospheres.

The molecule carbon monoxide (CO) is one key to detect the radial velocity of an exoplanet, since it exhibits a dense forest of deep absorption lines in a 
spectral band around 2.3~$\mu$m. Models predict CO to be one of the most 
abundant molecules in hot gas giant exoplanets \citep{2006ApJ...649.1048C,2007ApJS..168..140S}. To test those models, some papers have already been published reporting the 
detection or suggesting the presence of CO in the atmosphere of 
some transiting hot Jupiters \citep{2008Natur.456..767G,2009ApJ...690L.114S,2009ApJ...699..478D,2010ApJ...712L.139T}. All these measurements  were obtained with very low-resolution (R $<$ 40) HST NICMOS 
NIR spectra or broad Spitzer photometry beween 3.6 and 24 
$\mu$m, both via transmission (primary transit) or emission (secondary 
eclipse) observations. \cite{Sne10} 
reported the detection of CO  
via  the analysis of the transmission spectrum of the transiting planet HD~209458b 
using high-resolution spectroscopy between 2.30 and 2.33~$\mu$m. This allowed these authors to directly measure the RV of a transiting hot Jupiter for the very first time. We note that all those CO detections were done on transiting planets, but neither atmospheric chemicals nor RV shifts have yet been detected for  non-transiting exoplanets.    

Here we present the results of our attempt to detect the motion of the non-transiting planet $\tau$ Boo b via observations of its atmospheric CO spectrum in the NIR, by using a technique similar to \cite{Sne10}. The main goal of our study has been the measurement of the orbital inclination and the absolute mass of the planet, but as a by-product we can also confirm the presence of CO.

\section{The planetary system of $\tau$ Boo} \label{S2}

Tau Boo b is a massive hot Jupiter, orbiting
its host star every 3.31 days, and is estimated to have an atmospheric 
temperature above 1800~K following the formalism in \cite{2007ApJ...667L.191L} assuming zero albedo and no energy redistribution from day-to-night side. Table~\ref{tab:tauboo} summarizes the
  parameters of the planet and its very bright host star. We updated the orbital ephemerides by an RV-analysis of high-resolution UVES spectra of $\tau$ Boo published in \cite{rod10}

  \begin{table}
    \caption{Parameters of the star $\tau$~Boo and its planetary
  companion. Abbreviations for the references are: B06 = \cite{2006ApJ...646..505B} and references therein,
     B97~=~\cite{1997ApJ...474L.119B}, 
     G98~=~\cite{1998A&A...334..221G},
     H00~=~\cite{2000ApJ...531..415H},
      VF05~=~\cite{2005ApJS..159..141V},
     VV09~=~\cite{2009ApJ...694.1085V}.}             
    \label{tab:tauboo}      
    \begin{center}                         
    \begin{tabular}{l r l l }        
      \hline\hline                 
      Parameter & Value & Error &Ref. \\
      \hline                        
      Star: \\
      Spectral type & F7 IV-V  & & B97 \\ 
      $K~(mag)$        & 3.36 & 0.05 & VV09\\
      $m_{\star}~ (\rm{M_{\odot}})$ & 1.33 & 0.11 & VF05 \\
      $R_{\star}~(\rm{R_{\odot}})$ & 1.31 & 0.06 & G98 \\
      $T_{\rm eff}$~(K) & 6360 & 80 & G98 \\
      $P_{\rm{rot}}~({\rm d})$ & 3.31 &  & B97 \\
      $v~ \sin i~ (\rm{km~s^{-1}})$ & 14.9 & 0.5 & H00 \\
      Age (Gyr) & $1.3$  & 0.4 & VF05 \\
      \hline
      Planet:\\
      $m_{\rm{p}} \sin i~(\rm{M_{\rm{Jup}}})$  & 4.1 & 0.34 & B06 \\
      $a~ (\rm{AU})$ & 0.0481 & 0.0028 & B06 \\
      $e$ & 0.023 & 0.015 & B06 \\
      $K_{\star}~(\rm{km~s^{-1}})$ & 0.4611 & 0.0076 & B06 \\
      Orbital period $P_{\rm orb}$ (d) & 3.312458 & 0.00002  & new \\
      ${T_{\phi=0}}$ (JD) & 2~454~267.497  & 0.0122  & new \\
      \hline                                   
    \end{tabular}
    \end{center}
  \end{table}
%

The maximum possible RV semi-amplitude of the companion $\tau$~Boo~b is calculated to be $K_{\rm{p,max}}=157.3\pm~4.4~\rm{km~s^{-1}}$ via  
\begin{equation} \label{equ:doppler12}
    K_{\rm{p,max}} = \Big(\frac{2\pi G~m_\star}{P_{\rm orb}}\Big)^{1/3} ~,
  \end{equation} 
 where $G$ is the gravitational constant, $m_\star$ the stellar mass, and $P_{\rm orb}$ the orbital period. Due to the absence of transits
  in high-precision photometry (Henry et al. 2000), orbital inclinations larger than $i=83^{\circ}$ can be excluded. Baliunas et al.~(1997) found that the star $\tau$~Boo rotates rapidly with a period
 commensurate with the orbital period of the
planet, suggesting tidal locking. 
{\citet{cat07} and \citet{don08} carried out spectropolametric observations of $\tau$~Boo and measured differential rotation in the star, confirming that its rotation is synchronized with the orbital motion of the planet and suggesting an orbital inclination around 40$^\circ$.  \citet{2003MNRAS.344.1271L} and  Rodler et al.~(2010) attempted to measure starlight reflected from $\tau$~Boo~b, and found candidate features of marginal significance, indicating orbital inclinations of $i=37^\circ$ and $41^\circ$, respectively.}
Under the assumption of a tidal lock and that the stellar
equator and the orbital plane are co-aligned, Rodler et al.~(2010) predicted an orbital inclination of $i\approx 46^{\circ}$ and a planetary mass of  ${m_{\rm p}\approx 5.6~{\rm M}_{\rm Jup}}$. 

  \section{Observations and data reduction}
 We observed $\tau$ Boo during 5 hours on June 10, 2011, by using CRIRES \citep{2004SPIE.5492.1218K}, mounted on the VLT/UT1 on Cerro Paranal, Chile. We collected a total of 116 high-resolution spectra of our target plus 2 spectra of the A-type star HD~116160, which does not show any intrinsic absorption lines in the observed wavelength regime and therefore could be used for the calibration of our telluric model. The date of the observations was selected in such a way that the observations were carried out at orbital phases between $\phi=0.55$ and 0.61 ($\phi=0$ corresponds to the mid-transit position if $i\approx90^\circ$), when the 
 {hot day side} of the planet facing the star was largely visible and consequently appeared bright in the NIR. In addition, based on the ephemerides in Table~1 and the value of $K_\star$, we expected a variation of the RV of the companion up to 40~km~s$^{-1}$ during the course of the observations.
 
 The observations were carried out with the CRIRES standard setting of order 24 and a reference wavelength of $\lambda=2.3252~\mu$m. We used the 0.2''~slit to achieve a spectral resolution of $R=100,000$. We made use of the AO-system of CRIRES fo minimize slit losses. Observing conditions were good, with a seeing between 0.6'' and 1.8''. The observations were taken at two different nodding positions along the slit with the intention to remove the faint OH-lines in the sky background. We chose the integration times in such a way that the peak count rates per exposure were low ($\sim4000$ counts) to avoid non-linear sensitivity in the four detectors.

   The data were reduced with IRAF\footnote{IRAF Project Home Page:
http://iraf.noao.edu/}. { All four detectors showed a pattern aligned with the reading direction. This pattern consisted
of alternating rows or columns of larger and smaller intensities than the mean value and is called odd-even effect.
We applied a correction for the odd-even effect\footnote{An algorithm to correct this effect is provided at http://www.eso.org/sci/facilities/paranal/instruments/crires/tools/}, which was most present in the data recorded with detectors 1 and 4, for which the reading direction was perpendicular to the spectral dispersion.}
Since after this step strong noise features coming from those two detectors were still present in the data, we discarded those two detectors from any further analysis. The frames were taken in an A-A-B-B-B-B-A-A- sequence, where A and B denote the different nodding positions along the slit. For all frames as well as for each of each two remaining detectors, we combined the two consecutive frames, which had been taken at the same nodding position. This reduced the number of spectra to 58. Nodded images were
then subtracted to remove sky background and dark current. White-light spectra obtained with the same instrumental configuration in the afternoon before and after the observations were used to flat-field the data. Using the {\tt apall} task, we first identified and optimally centered the orders in the two individual nodding frames and traced these orders by adopting a second-order Legendre polynomial along the dispersion axis.
   We then extracted the one-dimensional spectra in detectors 2 and 3 and calculated a second-order polynomial wavelength solution by adopting as a reference system the telluric lines present in the spectra. The vacuum wavelengths of the telluric lines were identified using the 
   {line-by-line radiative transfer model (LBLRTM)}
    routine, which is based on the FASCODE algorithm \citep{1992JGR....9715761C}. The detectors 2 and 3 covered the wavelength regions 2.303 to 2.317~$\mu$m and 2.319 to 2.332~$\mu$m, respectively, with a pixel size corresponding to $\approx 1.6$~km~s$^{-1}$. We furthermore identified and removed obviously bad pixel in each spectrum coming from  defects of the detector.

   \section{Data analysis}
   \subsection{Overview}

   The observed target spectra were heavily contaminated by telluric lines, and to much less extent, by the rotationally broadened stellar absorption lines of the host star. A crucial step of the data analysis was the removal of the telluric and stellar spectra, and the search for the planetary spectrum in the residuals. In the following, we provide a description of the different data analysis steps. 

\subsection{Step 1: Determination of the instrumental profile}

To calculate the instrumental profile (IP) of the spectrograph for each target observation, we made use of the telluric contamination, which dominated our data, and of the telluric model spectra (see Step~2). In contrast to the large number of telluric lines, the stellar spectrum of $\tau$ Boo exhibits very few absorption features in the observed wavelength regime, which appeared rotationally broadened due to the high $v \sin i$ (Table~1). Those regions of stellar absorption features were excluded from the determination of the instrumental profile. We normalized the spectra in such a way that the flux in the telluric/pseudo-stellar continuum was at one, and stored the photon noise error per spectral pixel. We then determined the IP as the sum of seven Gaussian profiles in a similar way as described in \cite{1995PASP..107..966V}: Around a central Gaussian we grouped three satellite Gaussians on each side of it, which allowed us to account for asymmetries in the IP. Free parameters were the width of the central Gaussian, plus the amplitudes of the six satellite Gaussians, while the positions and the widths of those satellites were fixed and set {\it a priori} in a way that their half-widths overlapped. By employing a Brent algorithm, we finally determined the free parameters and convolved the telluric model spectrum with the IP. We found that for all spectra, the shape of the IP remained constant. However, the width of the IP was slightly different at each nodding position and was also changing during the course of the night.

\subsection{Step 2: Telluric model}

For the calculation of the atmospheric transmission spectrum, we used the LBLRTM code, which is available as a Fortran source code\footnote{Source code and manuals are available under {\tt http://rtweb.aer.com/lblrtm\_description.html}}. As molecular database we adopted HITRAN \citep{2005JQSRT..96..139R}, which contains the 42 most prominent molecules and isotopes present in the atmosphere of the Earth. Following the approach presented by \cite{2010A&A...524A..11S}, we created a high-resolution theoretical vacuum-wavelength telluric spectrum for each observed spectrum by accounting for the air mass of the star as well as the weather conditions (water vapour density column, temperature and pressure profiles) during the observations. We retrieved the weather information from the Global Data Assimilation System (GDAS). GDAS models are available in 3 hour intervals for any location around the globe\footnote{GDAS webpage: {\tt http://ready.arl.noaa.gov/READYamet.php}}.

We first calibrated the line depths of the telluric lines by using the A-star observations, which had been taken at the beginning of the night. For five water and methane lines, the line depths were systematically underestimated in the HITRAN model; for these lines we consequently updated the abundances in our telluric model. In addition to that, the amount of water vapor in the atmosphere constantly changed during the observations, which resulted in variations of the line depths of the water lines in the observed spectra. Consequently, for each observed target spectrum, we created a set of two telluric model spectra, one for water and one for the other molecules, most notably CH$_4$. We then globally scaled the line depths for those two models for each target observation. Finally, we multiplied the two different high resolution telluric spectra, convolved it with the IP to create the telluric model.

\subsection{Step 3: Stellar model}
The few stellar absorption lines remained in the data after the removal of the telluric lines. We attempted to subtract the stellar spectra from the data using models, but found that existing models (MARCS, \citealp{2008A&A...486..951G}; PHOENIX, \citealp{1997ApJ...483..390H}) do not reproduce the observed stellar spectrum of $\tau$ Boo. The best alternative was to create a high signal-to-noise ratio template by co-adding all the observed telluric-free spectra.
Due to the expected large orbital motion of the planet, contributions to this template spectrum coming from the planet, which is estimated to be a thousand times fainter than the star, were smeared out to a large extent.
Before combining the individual stellar spectra to one template spectrum, we applied a Savitzky-Golay smoothing algorithm \citep{pre92} to each of them, which ensured that the broad stellar lines remained in the spectra, while the sharp and unseen planetary lines were smoothed out. In the final step, we subtracted the stellar template spectrum from all telluric-free target spectra.

\subsection{Step 4: Searching for the CO features}

The mean RV of the host star $\tau$ Boo with respect to the barycenter of the Solar system is $-15.8\pm0.5$~km~s$^{-1}$ (blue-shifted; 	
\citealp{2000A&AS..142..217B}). The barycentric RV of the Earth with respect to the sky position of the target at the time of observations was between -22.8~km~s$^{-1}$ and -23.4~km~s$^{-1}$. This means that during the times of our measurements, the stellar spectrum of $\tau$ Boo was observed with a red-shift of about 8~km~s$^{-1}$. That red-shift was subtracted from the resultant RVs of the planetary signal, as described below.

After the removal of the stellar and telluric absorption lines, we corrected the residual spectra for trends originating from intra-pixel variations. We then cross-correlated the telluric-free and stellar-free target residual spectra with a CO-model spectrum (Fig.~1), which was calculated with the PHOENIX code \citep{1997ApJ...483..390H,2008ApJ...675L.105H,2011A&A...529A..44W} for a brown dwarf having a temperature of $T_{\rm eff}=1800$~K and Solar metallicity. To this end, we shifted the CO-model spectrum for each residual spectrum according to the instantaneous radial velocity $V_{\rm{p}}(\phi,i)$ with respect to the star, which depends on both the orbital phase $\phi$, which was a priori known, and the unknown orbital inclination $i$ as:
  \begin{equation} \label{equ:doppler1}
    V_{\rm{p}} = K_{\rm{p,max}} \sin 2\pi\phi \sin i = K_{\rm p} \sin 2\pi\phi ~.
  \end{equation} 
 For all residual spectra, we then calculated the correlation value for different RV semi-amplitudes $K_{\rm p}$ of the planetary signal and finally determined the cross-correlation function (CCF) with respect to $K_{\rm p}$.

   \subsection{Step 5: Confidence level}

   The confidence level of the strongest peak of the CCF was determined by employing a
   bootstrap randomisation method (e.g. \citealp{1997A&A...320..831K}): We assigned random values of the orbital phases to the observed spectra, thereby creating $N$ different data sets.
   Any signal present in the original data was then scrambled in these
   artificial data sets.
   For all these randomized data sets, we again evaluated the model for the
   free parameter $K_{\rm p}$ and located the best fit with its specific CCF-peak value. 
   
The confidence level of the CCF peak was estimated to be $\approx 1-m/N$, where $m$ is the number of the best fit models  having a CCF-peak value larger or equal than the maximum of the CCF found in the original, unscrambled
   data sets. Notice that we consider a detection to have a confidence of $\ge 0.9987$, which translates to $\ge 3\sigma$.

\section{Results and discussion}
As illustrated in Figures 2 and 3, the results of our cross-correlation analysis using a CO-model spectrum
reveal Doppler velocity shifts consistent with a maximum RV semi-amplitude for the planet of $K_{\rm p} = 115 \pm 11~$km~s$^{-1}$. { The $1\sigma$-errorbar contains both the actual error of the velocity measurement, which was determined via bootstrap resampling \citep{bar92}, as well as the shift in velocity due to the uncertainties in the orbital solution.} In addition to that, our bootstrap randomisation analysis revealed that this signal is significant at the $3.4\sigma$ confidence level, with 2 false positives in 3000 trials. 

\begin{figure}[t!]
   \centering
   \includegraphics[angle=-90,width=8.8cm]{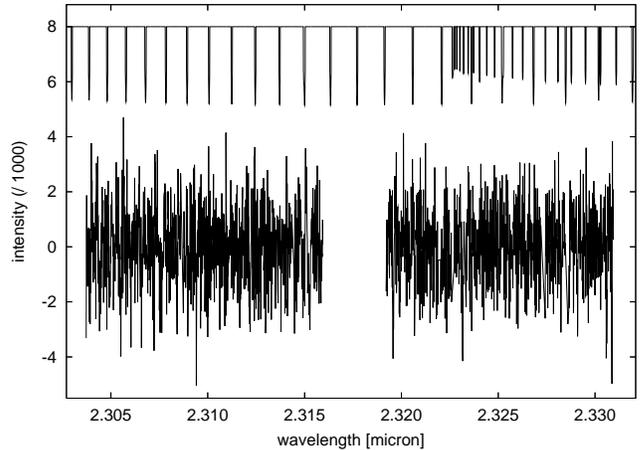}
      \caption{Bottom: Residual spectrum after subtraction of the stellar and telluric absorption lines from the data.   The upper spectrum depicts the model of the dense forest of carbon-monoxide (CO) absorption lines. For clarity, the line depths in the CO model are shown for a planet-to-star flux ratio of $2\times10^{-3}$, and the spectrum is shifted up by 3 units.}
          \label{plot01}
   \end{figure}

\begin{figure}[t!]
   \centering
   \includegraphics[angle=-90,width=9cm]{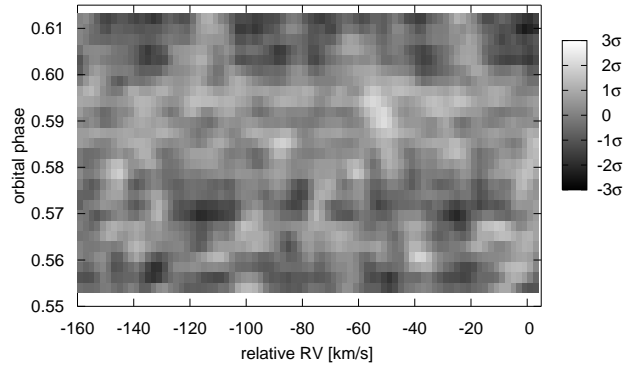}
      \caption{CO absorption in the planet atmosphere of $\tau$~Boo~b. The individual cross-correlation functions (CCFs) of the 58 residual spectra with the CO model spectrum in the rest-frame of the star are shown. During the course of the observations, the orbital motion of the planet produces a RV-shift starting at about -35~km~s$^{-1}$ and ending at 65~km~s$^{-1}$, respectively for  orbital phases of 0.55  and 0.61. The linear grey-scales indicate the strength of the cross-correlation signal (bright means absorption, dark means emission).}
          \label{plot02}
   \end{figure}

\begin{figure}[t!]
   \centering
   \includegraphics[angle=-90,width=8.7cm]{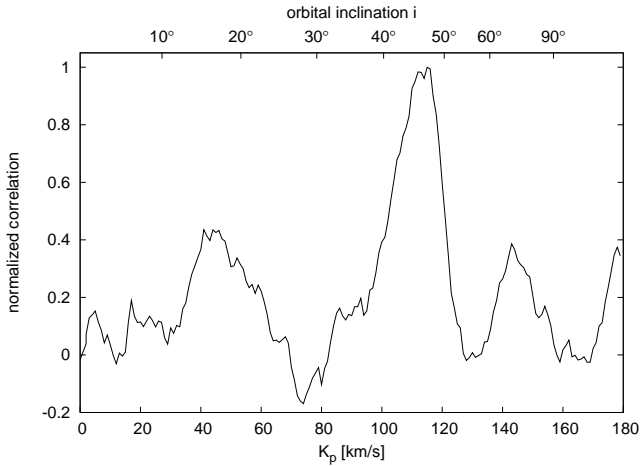}
      \caption{The CO signal co-added for all spectra. The cross-correlation functions (CCF) from all 58 residual spectra were combined assuming different RV semi-amplitudes of $\tau$~Boo~b. The peak of the CCF occurs at a RV semi-amplitude of $115\pm 11$~km~s$^{-1}$, which corresponds to an orbital inclination of $i=47^{+7}_{-6}$~degrees. The peak of the CCF is significant at the $3.4\sigma$ confidence level.}
          \label{plot03}
   \end{figure}

The measured RV semi-amplitude corresponds to an orbital inclination for the system of $i = 47^{+7}_{-6}$~degrees. Adopting the measured orbital inclination, the value of the stellar mass in Table1 and Kepler Laws of planetary motion, we finally are able to derive an absolute mass for the planet of $m_{\rm p} = 5.6 \pm  0.7~{\rm M}_{\rm Jup}$. This value clearly confirms the planet hypothesis for the hot Jupiter $\tau$~Boo~b. 

Given the precision of our data, possible residuals of the planetary spectrum in the stellar template, the lack of detailed atmospheric temperature-pressure profile models for this planet, and uncertainties in the possible level of masking between CO and CH$_4$ (both molecules are expected to co-exist in $\tau$ Boo, e.g. \citealp{1999ApJ...512..843B}), we cannot provide a reliable estimation of the CO abundance in the atmosphere of $\tau$ Boo b. However, we can place a lower limit to the planet-to-star flux ratio in the observed wavelength range based on the atmospheric model we adopted. To this end, we created data sets which were based on our residual spectra. For each of these data sets, we injected an artificial planetary CO signal having a RV semi-amplitude of $K_{\rm p} = 115$~km~s$^{-1}$ with selected planet-to-star flux ratios from  $10^{-2}$ down to $10^{-4}$. To avoid an overlapping of the CCF peaks with the real signal, we red-shifted the spectrum of the injected planet by 50~km~s$^{-1}$. For each data set, we then determined that peak of the CCF apart from the one of real signal and determined its confidence level. We find that with our data set and our CO-model, we are sensitive to detect a planet with planet-to-star flux ratio larger than $\approx7\times10^{-4}$ at the 3$\sigma$ confidence level.

We emphasize that this measurement of the exact mass of $\tau$~Boo~b represents the first successful determination of  the mass of a non-transiting planet by means of high-resolution spectroscopy. 
This technique has therefore the potential of providing direct masses and estimation of the atmospheric composition of non-transiting exoplanets in the near-future, in particular as new facilities like E-ELT become available.


\acknowledgments
Based on observations made with ESO Telescopes at the Paranal Observatory under programme ID 087.C-0407. This work has been partially supported by NSF's grant AST-0908278.
We are very grateful to Andreas Seifahrt for his help with LBLRTM and HITRAN, to Martin K\"urster for input for data analysis strategies and statistics, and to John Barnes for discussions about the search for thermal emission from hot Jupiters. We would further like to thank Ignas Snellen and Matteo Brogi for their help, and to the anonymous referee for very constructive suggestions which significantly improved the manuscript. 


\end{document}